\documentclass[aps,prc,superscriptaddress,showpacs,onecolumn]{revtex4-1}
 \usepackage{amsmath}  
 \usepackage{makeidx}
 \usepackage{graphicx}
 \usepackage{amsfonts}
 \usepackage[ansinew]{inputenc}
 \usepackage[usenames,dvipsnames]{pstricks}
 \usepackage{subfigure}
 \usepackage{epstopdf}
\usepackage{epsfig}
 \usepackage{mathrsfs}

              \makeindex

\begin{document}
\title{Alpha Decay Width of $^{212}$Po from a quartetting wave function approach}
\author{Chang Xu}
\email{cxu@nju.edu.cn}
\affiliation{Department of Physics and Institute of Acoustics, Nanjing University, Nanjing 210093, China}
\author{Zhongzhou Ren}
\email{zren@nju.edu.cn}
\affiliation{Department of Physics and Institute of Acoustics, Nanjing University, Nanjing 210093, China}
\affiliation{Center of Theoretical Nuclear Physics, National Laboratory of Heavy-Ion Accelerator, Lanzhou 730000, China}
\author{G. R\"{o}pke}
\email{gerd.roepke@uni-rostock.de}
\affiliation {Institut f\"{u}r Physik, Universit\"{a}t Rostock, D-18051 Rostock, Germany}
\author{P. Schuck}
\email{schuck@ipno.in2p3.fr}
\affiliation{Institut de Physique Nucl\'{e}aire, Universit\'e Paris-Sud, IN2P3-CNRS, UMR 8608, F-91406, Orsay, France}
\affiliation{Laboratoire de Physique et Mod\'elisation des Milieux Condens\'es, CNRS- UMR 5493,
F-38042 Grenoble Cedex 9, France}
\author{Y. Funaki}
\affiliation{RIKEN Nishina Center,
Wako 351-0198, Japan}
\author{H. Horiuchi}
\affiliation {Research Center for Nuclear Physics (RCNP), Osaka University, Osaka  567-0047, Japan}
\affiliation {International Institute for Advanced Studies, Kizugawa 619-0225,  Japan}
\author{A. Tohsaki}
\affiliation{Research Center for Nuclear Physics (RCNP), Osaka University, Osaka 567-0047, Japan}
\author{T. Yamada}
\affiliation{Laboratory of Physics, Kanto Gakuin University, Yokohama 236-8501, Japan}
\author{Bo Zhou}
\affiliation{Research Center for Nuclear Physics (RCNP), Osaka University, Osaka 567-0047, Japan}


\date{\today}

\begin{abstract}
A microscopic calculation of $\alpha$-cluster preformation probability and $\alpha$ decay width in the
typical $\alpha$ emitter $^{212}$Po is presented.
Results are obtained by improving a recent approach to describe $\alpha$
preformation in $^{212}$Po [Phys. Rev. C {\bf 90}, 034304 (2014)]
implementing four-nucleon correlations (quartetting).
Using the actually measured density distribution of the  $^{208}$ Pb
core, the calculated alpha decay width of $^{212}$Po agrees fairly
well with the measured one.
\end{abstract}


\maketitle

The radioactive $\alpha$ decay is a frequent phenomenon  in
nuclear physics, in particular near the doubly magic nuclei
$^{100}$Sn and $^{208}$Pb, and the superheavies where $\alpha$
decay competes with spontaneous fission (for a recent discussion see \cite{DLW}
and refs. given there). Whereas the tunneling of
an $\alpha$ particle across the Coulomb barrier is well described
in quantum physics, the problem in understanding the $\alpha$
decay within a microscopic approach is the preformation of the
$\alpha$ cluster in the decaying nucleus.

The formation of
$\alpha$-like correlations in nuclear systems has been
investigated recently. In particular, in light, low density states
of selfconjugate nuclei ($^8$Be, $^{12}$C, $^{16}$O, $^{20}$Ne,
etc.) four-nucleon correlations have been identified within the
THSR (Tohsaki-Horiuchi-Schuck-R\"{o}pke) approach \cite{THSR}, but also with
other theories like Resonating Group Method (RGM) \cite{RGM}, Brink-GCM
(Generator Coordinate Method) \cite{GCM}, Fermion Molecular Dynamics (FMD) \cite{FMD},
Antisymmetrised Molecular Dynamics (AMD) approaches \cite{AMD} going beyond
the mean-field approximation. The main message is that
well-defined clusters are formed only in regions where the density
of nuclear matter is low. Therefore, it is of interest to
investigate $\alpha$-like correlations also in the outer tails of
the density of a nucleus, and $\alpha$ preformation is discussed as a
surface effect confined to the region where the nucleon density is
comparable or below 1/5 of saturation density $n_{\rm sat}=0.16$
fm$^{-3}$.

A typical example is $^{212}$Po which is an $\alpha$ emitter with
half-life 0.299 $\mu$s and decay energy $Q_\alpha = 8954.13$ keV. It is
spherical, doubly magic, and has only one decay channel. Several approaches have been made to calculate the $\alpha$ decay width of
$^{212}$Po within a microscopic approach, see Ref. \cite{Varga} and refs. therein.
Furthermore a quartetting wave function approach has been worked
out recently \cite{Po}. Performing various approximations,
exploratory calculations resulted in a preformation factor of
about 0.37.

Here, we are interested in the $\alpha$ decay width of $^{212}$Po.
The transition probability for the $\alpha$ decay  $W=P_{\alpha}
\nu {\cal T}$ is given as product of the preformation probability
$P_{\alpha}$, the frequency (pre-exponential factor) $\nu$, and
the exponential factor ${\cal T}$. In the present work, we improve
the exploratory calculation performed in \cite{Po} replacing
simple expressions for the density and the potentials by more
realistic ones. We refine the calculation of the quartetting state
using recently measured density profiles for the  $^{208}$Pb core
\cite{Tarbert2014}. Furthermore, we improve the mean-field
potential using the M3Y double-folding potential, see
\cite{Misicu2007}, instead of the Woods-Saxon potential. To
evaluate the $\alpha$ decay width, we use the approach of Gurvitz
\cite{Gurvitz1988} to estimate the preexponential factor.
In addition to the preformation factor and the binding energy, results for the half-life will be given.

{\it Preformation probability.} An effective $\alpha$ particle equation has been derived recently
\cite{Po} for cases where an $\alpha$ particle is bound to a
doubly magic nucleus. As an example, $^{212}$Po has been
considered which decays into a $^{208}$Pb core and an $\alpha$ particle.
Neglecting recoil effects, we assume that the core nucleus is fixed at
${\bf r} = 0$. The core nucleons are distributed with the baryon density $n_B(r)$ and produce
a mean field $V_\tau^{\rm mf}(r)$ acting on the two neutrons ($\tau = n$)
and two protons ($\tau = p$) moving on top of the lead core.
In the present work, we will not give a microscopic description
of the core nucleons (e.g., Thomas-Fermi or shell model calculations)
but consider both $n_B(r)$ and $V_\tau^{\rm mf}(r)$ as phenomenological input.
Of interest is the wave function of the four nucleons on top of the core nucleus
which can form an $\alpha$-like cluster.

The four-nucleon wave function (quartetting state) $\Psi({\bf
R},{\bf s}_j)=\varphi^{\text{intr}}({\bf s}_j,{\bf R})\,\Phi({\bf
R})$ is subdivided in a unique way in the (normalized) center of mass (c.m.)
part $\Phi({\bf R})$ depending only on the c.m. coordinate $\bf R$,
and the intrinsic part $\varphi^{\text{intr}}({\bf s}_j,{\bf R})$
which depends, in addition, on the relative coordinates ${\bf
s}_j$ (for instance, Jacobi-Moshinsky coordinates)
\cite{Po}. The respective c.m. and intrinsic Schr\"odinger
equations are coupled by contributions containing the expression
$\nabla_R \varphi^{\text{intr}}({\bf s}_j,{\bf R})$ which will be
neglected in the present work. In contrast to homogeneous matter
where this expression disappears, in finite nuclear systems such
as $^{212}$Po this gradient term will give a contribution to the
wave equations for $\Phi({\bf R})$ as well as for
$\varphi^{\text{intr}}({\bf s}_j,{\bf R})$. Up to now, there are
no investigations of such gradient terms.

The intrinsic wave equation describes in the zero density limit
the formation of an $\alpha$ particle with binding energy $B_\alpha= 28.3$
MeV. For homogeneous matter, the binding energy will be reduced
because of Pauli blocking. In the zero temperature case considered
here, the shift of the binding energy is determined by the baryon
density $n_B=n_n+n_p$, i.e. the sum of the neutron density $n_n$
and the proton density $n_p$. Furthermore, Pauli blocking depends on the asymmetry given by
the proton fraction $n_p/n_B$ and the c.m. momentum ${\bf P}$ of the
$\alpha$ particle. Neglecting the weak dependence on the
asymmetry, for ${\bf P}=0$ the density dependence of the Pauli
blocking term
$W^{\rm Pauli}(n_B)=4515.9\, n_B -100935\, n_B^2+1202538\, n_B^3$
was found in \cite{Po}, Eq. (45), as a power expansion with respect to $n_B$.
In particular, the bound state is dissolved and merges with the continuum
of scattering states at the Mott density $n_B^{\rm Mott}= 0.02917$ fm$^{-3}$.

The intrinsic wave function remains nearly  $\alpha$-particle like up
to the Mott density (a small change of the width parameter $b$ of
the four-nucleon bound state is shown in Fig. 2 of Ref.
\cite{Po}), but becomes a product of free nucleon wave functions
(more precisely the product of scattering states) above the Mott
density. This behavior of the intrinsic wave function will be used
below when the preformation probability for the  $\alpha$ particle
is calculated. Below the Mott density the intrinsic part of the
quartetting wave function has a large overlap with the intrinsic
wave function of the free $\alpha$ particle.
In the region where the $\alpha$-like cluster
penetrates the core nucleus, the intrinsic bound state wave
function transforms at the critical density $n_B^{\rm Mott}$ into an unbound
four-nucleon shell model state.

In the case of $^{212}$Po considered here, an $\alpha$ particle is
moving on top of the doubly magic $^{208}$Pb core. The tails of
the density distribution of the Pb core where the baryon density
is below the Mott density $n_B^{\rm Mott}$, is relevant for
the formation of  $\alpha$-like four-nucleon correlations. Simply
spoken, the $\alpha$ particle can exist only at the surface of the
heavy nucleus. This peculiarity has been considered since a long
time for the qualitative discussion of the preformation of
$\alpha$ particles in heavy nuclei \cite{DLW,Xu,Denisov}. It
has recently also been discussed in connection with the neutron
skin thickness of heavy nuclei with $\alpha$-particle correlations
\cite{Typel2014}.

Improving simple estimations (Thomas-Fermi
model as well as the Shlomo parametrization of the density) for the baryon density considered in
\cite{Po}, we use the empirical results obtained recently
\cite{Tarbert2014} which are parametrized by Fermi functions. With
the neutron density $n_{n}( r)=0.093776/\{1+\exp[(r-6.7 {\rm
fm})/0.55 {\rm fm}]\} {\rm fm}^{-3}$ and the proton density
$n_{p}( r)=0.062895/\{1+\exp[(r-6.68 {\rm fm})/0.447 {\rm fm}]\}
{\rm fm}^{-3}$,  the Mott density $n_B^{\rm Mott}= 0.02917$ fm$^{-3}$ occurs
at $r_{\rm cluster} =7.4383$ fm, $n_B(r_{\rm cluster})=n_B^{\rm Mott}$.
This means that $\alpha$-like
clusters can exist only at distances $r > r_{\rm cluster}$, for
smaller values of $r$ the intrinsic wave function is characterized
by the uncorrelated motion.Note that this transfer of results
obtained for homogeneous matter to finite nuclei is based on a
local density approach. In contrast to the weakly  bound
di-nucleon cluster, the $\alpha$ particles are more compact so
that a local-density approach seems to be better founded. However,
the Pauli blocking term is non-local. As shown in \cite{Po}, the
local density approach can be improved systematically. It is
expected that non-local  interaction terms and gradient terms will
make the sudden transition  at $r_{\rm cluster}$ from the
intrinsic $\alpha$-like cluster wave function to an uncorrelated four-nucleon
wave function more smooth.

Our main attention is focussed on the c.m. motion  $\Phi({\bf R})$
of the four-nucleon wave function (quartetting state of four nucleons
$n_\uparrow, n_\downarrow, p_\uparrow, p_\downarrow$). Because
the lead core nucleus is very heavy, we replace the c.m.
coordinate $\bf R$ by the distance $r$ from the center of the
$^{208}$Pb core. The corresponding Schr\"odinger equation contains
the kinetic part $-\hbar^2 \nabla_r^2/8 m$ as well as the
potential part $W({\bf r},{\bf r}')$ which, in general, is
non-local but can be approximated by an effective c.m. potential
$W(r)$. The effective c.m. potential consists of two
contributions, the intrinsic part $W^{\rm intr}( r)=E_\alpha^{(0)} + W^{\rm Pauli}(r)$ and the external
part $W^{\rm ext}(r)$ which is determined by the mean-field
interactions.

\begin{figure}[h]
   \includegraphics[width=0.5\textwidth]{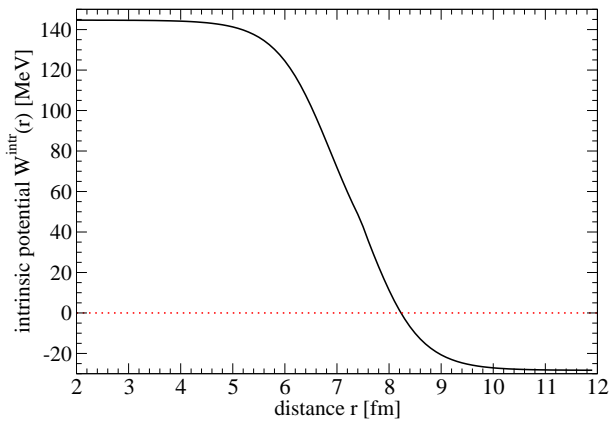}
    \caption{Intrinsic part $W^{\rm intr}( {\bf r})$ of the effective c.m. potential $W( {\bf r})$. The empirical density distribution \cite{Tarbert2014} for the  $^{208}$Pb core has been used.
    The four-nucleon Fermi energy for $r <  r_{\rm cluster}$ is taken in Thomas--Fermi approximation}
 \label{Fig:intrW}
 \end{figure}

The intrinsic part $W^{\rm intr}(r)$ approaches for large
$r$ the bound state energy $E_\alpha^{(0)} =- B_\alpha= -28.3$ MeV of the $\alpha$ particle.
In addition, it contains the Pauli
blocking effects $W^{\rm Pauli}(r)$ given above \cite{Po}.
Since the distance from the center of the lead
core is now denoted by $r$, we have for $r >  r_{\rm cluster}$ the
shift of the binding energy of the $\alpha$-like cluster.
Here, the Pauli blocking part has the form ${ W}^{\rm Pauli}[n_B(r)]$ given above.
For $r < r_{\rm cluster}$, the density of the core nucleus is larger than
the Mott density so that no bound state is formed. As lowest
energy state, the four nucleons of the quartetting state are added
at the edge of the continuum states that is given by the chemical potential.
In the case of the Thomas-Fermi model, not accounting
for an external potential, the chemical potential coincides with
the sum of the four constituting Fermi energies. For illustration,
the intrinsic part $W^{\rm intr}( {\bf r})$ in Thomas-Fermi
approximation, based on the empirical density distribution, is
shown in Fig. \ref{Fig:intrW}.
The repulsive contribution of the Pauli exclusion principle is
clearly seen.

The external part $W^{\rm ext}( {\bf r})$ is given by the mean
field of the surrounding matter acting on the four-nucleon system.
It includes the strong nucleon-nucleon interaction as well as the
Coulomb interaction. According to Eq. (51) in \cite{Po} it is
given by a double-folding potential using the intrinsic
$\alpha$-like cluster wave function. For $r >  r_{\rm cluster}$
the simple Woods-Saxon potential used in \cite{Po} is improved in
the present work using the M3Y double-folding potential
\cite{M3YReview}. This M3Y potential contains in addition to the
Coulomb interaction the direct nucleon-nucleon interaction $V_N(
r)$ and the exchange terms $V_{\rm ex}( r)+V_{\rm Pauli}( r)$
\cite{M3YReview}.

 \begin{figure}[th]
    \includegraphics[width=0.5\textwidth]{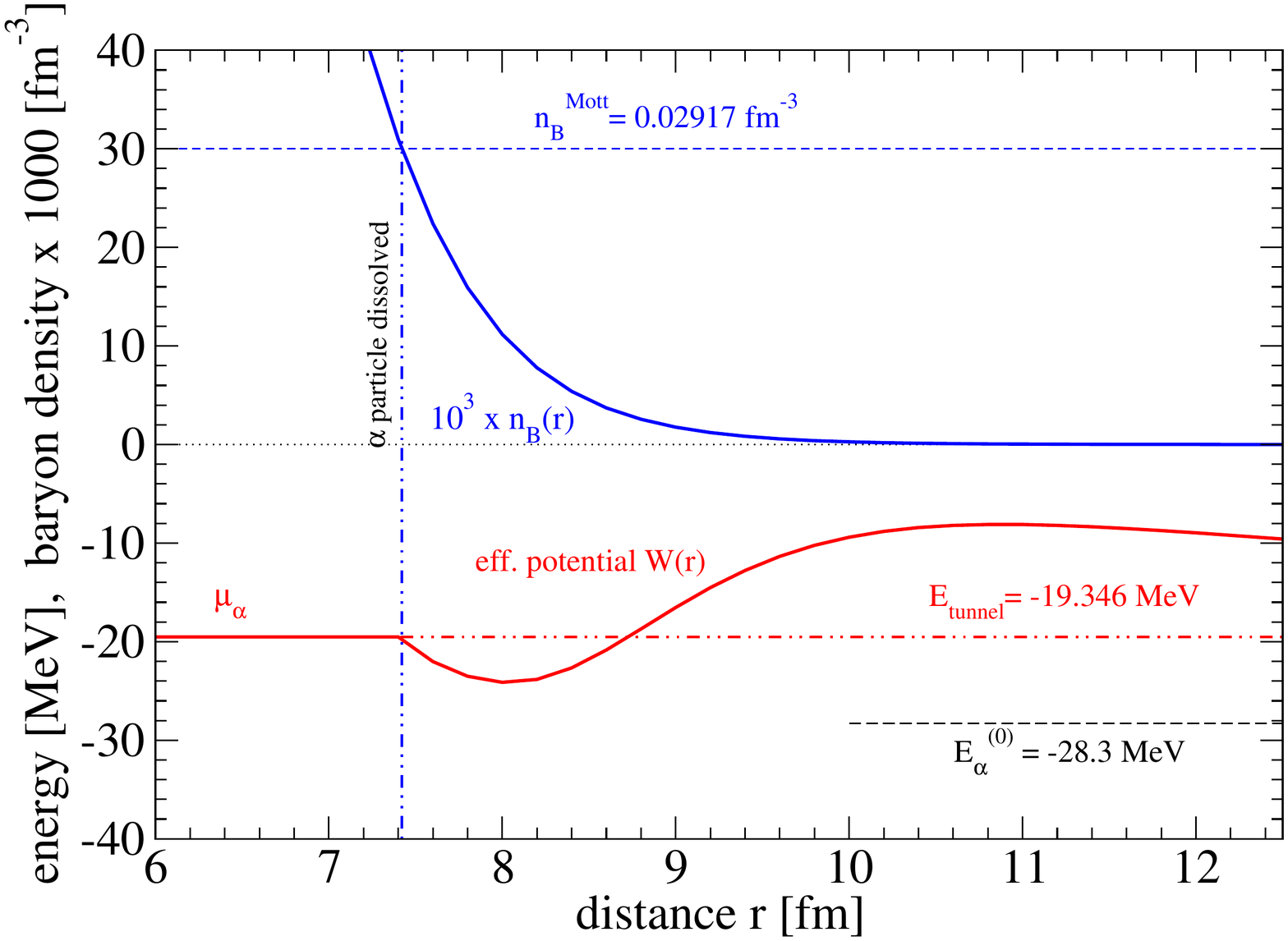}
    \caption{Effective c.m. potential $W( r)$, potential A.
The empirical  baryon density distribution \cite{Tarbert2014} for the  $^{208}$Pb core is also shown.
The chemical potential $\mu_\alpha$ coincides with the binding energy $E_{\rm tunnel}$. }
 \label{Fig:2a}
 \end{figure}

The Coulomb interaction is calculated as a  double-folding potential
using the proton density $n_p(r )$ of the $^{208}$Pb core given above
and a Gaussian density distribution for the $\alpha$ cluster,
with the charge r.m.s. radius 1.67 fm. The direct
nucleon-nucleon interaction is
obtained by folding the measured nucleon density distribution of the
$^{208}$Pb core $n_B( r)$ and the Gaussian density distribution for the $\alpha$ cluster
(point r.m.s. radius 1.44 fm)  with a parameterized  nucleon-nucleon
effective interaction $v(s)=c \exp(-4s)/(4s)-d \exp(-2.5s)/(2.5s)$
describing a short-range repulsion ($c$) and a long-range attraction ($d$),
$s$ denotes the nucleon-nucleon distance.

We will not perform a self-consistent calculation of the core nucleus here, but
consider the potential $W(r)$ in some approximations.
For comparison, three sets of c.o.m.
potentials from the double-folding procedure are discussed here: Potential A, potential B,
and potential C. For $r >  r_{\rm cluster}$, the  corresponding parameter values for $c,d$ of the
direct term $V_N( r)$ are given in Tab. \ref{Tab:1}.
For the exchange terms we use in case A and B the
Pauli blocking  ${ W}^{\rm Pauli}[n_B(r )]$ obtained from the microscopic approach,
in case C the M3Y parametrization for the exchange terms are given below.
Potential A is considered to explain the physics behind our approach.

In principle, the nucleon mean field should
reproduce  the empirical densities of the $^{208}$Pb core. For $r
<  r_{\rm cluster}$  a local-density (Thomas-Fermi) approach will give a
constant chemical potential $\mu_{4}$ which is the sum of the
mean-field potential and the Fermi energy of the four nucleons,
$\mu_{4}=W^{\rm ext}( {\bf r})+2E_{{\rm F},n}(n_n)+2E_{{\rm F},p}(n_p)$ with
$E_{{\rm F},\tau}(n_\tau)=(\hbar^2/2 m_\tau) (3 \pi^2 n_\tau)^{2/3}$.
This quantity is not depending on position. Additional four nucleons must be
introduced at the value $\mu_4$.
We consider this property as valid for any local-density approach, the continuum edge for adding
quasiparticles to the core nucleus is given by the chemical potential, not depending on position.
In a rigorous Thomas--Fermi approach for the core nucleus, this chemical potential coincides with the bound state
energy $E_{\rm tunnel}$ of the four-nucleon cluster, $E_{\rm tunnel}=\mu_4$.
For $r <  r_{\rm cluster}$, the effective c.m. potential $W(r)$ describes the edge of the
four-nucleon continuum where the nucleons can penetrate into the core nucleus.
Note that we withdraw this relation for shell model calculations where all states below the Fermi energy
are occupied, but the next states (we consider the states above the Fermi energy as "continuum states"
with respect to the intrinsic four-nucleon motion) are separated by a gap so that $E_{\rm tunnel}>\mu_4$.
We come back to this issue below in the Discussion.

Potential A is designed according to this simple local-density approach. The two parameter
values $c_{\rm A}, d_{\rm A}$ are determined by the conditions $\mu_4=E_{\rm tunnel}=-19.346$ MeV,
see Fig. \ref{Fig:2a}. The tunneling energy is identical with the energy at which the four nucleons
are added to the core nucleus. The total c.m. potential is continuous at $r=  r_{\rm cluster}$
and is constant for $r<  r_{\rm cluster}$, where the effective c.m. potential is
$W( { r}) =\mu_{4}$. The corresponding values for the preformation factor
and the decay half-life are given in Tab. \ref{Tab:1}.

In a better approximation, the simple local-density (Thomas-Fermi) approach
for the $^{208}$Pb core nucleus has to be replaced by a shell model calculation. Then, the
single-particle states are occupied up to the Fermi energy, and additional nucleons
are introduced on higher energy levels according to the discrete
structure of the single energy level spectrum. The condition   $E_{\rm tunnel}=\mu_4$ is withdrawn.
If the shell model is appropriate for the core nucleus, with the Fermi energy at  $\mu_4$,
additional nucleons have a somewhat higher energy,  $E_{\rm tunnel}>\mu_4$. (Note that in the opposite case
a shell-model approach becomes unstable against the formation of clusters.)

\begin{figure}[th]
\includegraphics[width=0.4\textwidth]{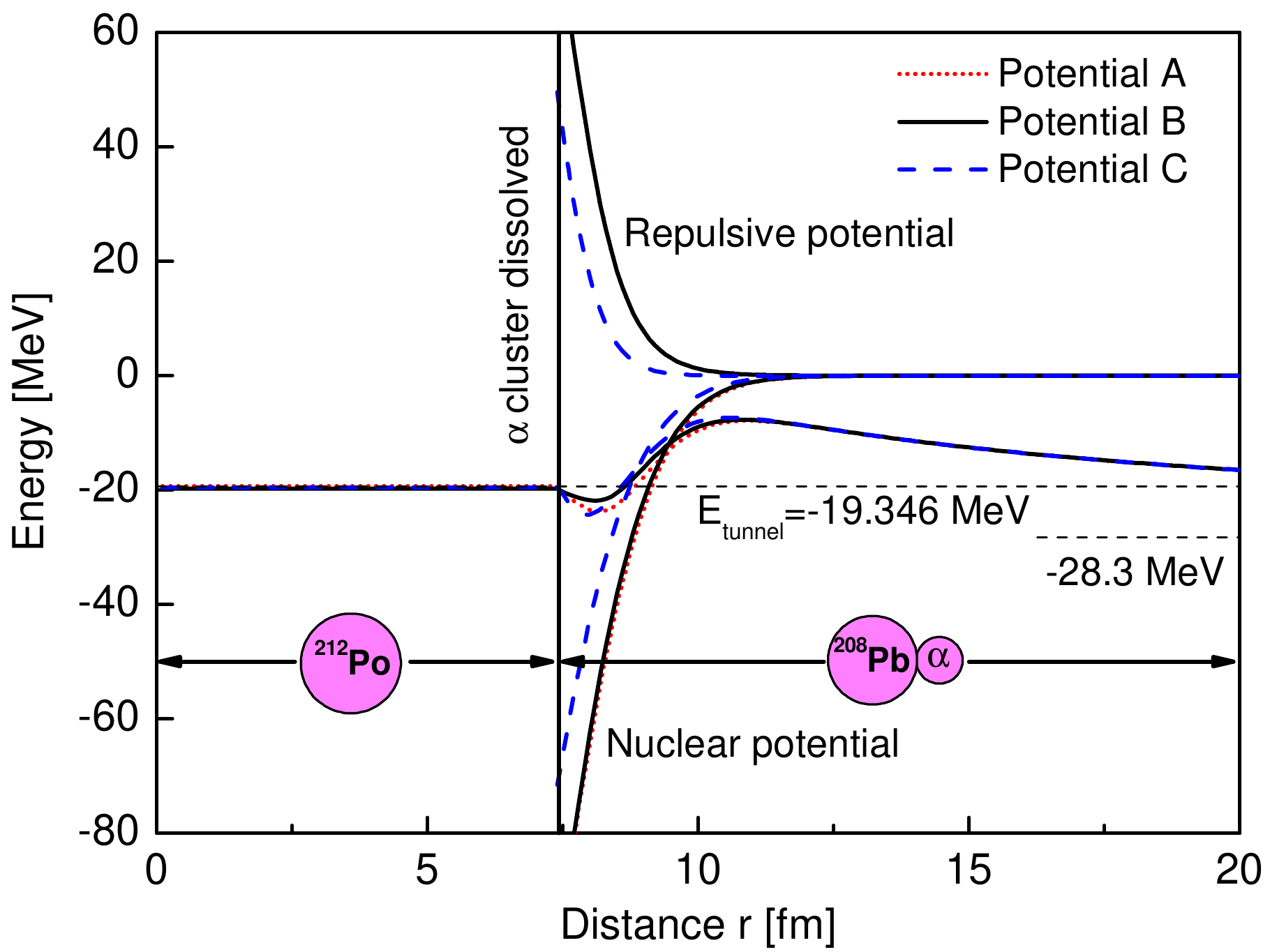}
\caption{Effective c.o.m. potential $W(r)$ for the
$\alpha$-decay of $^{212}$Po. Both versions B, C give the empirical
bound state energy $E_{\rm tunnel}=-19.346$ MeV, see Tab. I.  The repulsive
potential is given by the Pauli blocking term (Potential B) or the exchange
terms (Potential C). The total c.m. potential $W(r)$ contains, in addition,
the Coulomb part and the bound state energy $E_\alpha^{(0)}=-28.3$ MeV. The chemical potential $\mu_4$
is slightly deeper than the bound state energy $E_{\rm tunnel}=-19.346$ MeV, see Tab. I. }
 \label{Fig:1}
 \end{figure}

Potential B is designed without the condition $\mu_4=E_{\rm tunnel}$. The two parameter
values $c_{\rm B}, d_{\rm B}$ are determined by the two empirical values bound state energy
 $E_{\rm tunnel}=-19.346$ MeV and the half-life$T_{1/2}=2.99 \times  10^{-7}$ s for $^{212}$Po.
 The corresponding values are given in Tab. \ref{Tab:1}. As expected, the bound state energy
 is above the value $W( { r})=\mu_4$ for the c.m. potential at $r<  r_{\rm cluster}$. A plot of the
 c.m. potential as well as the different contributions is shown in Fig. \ref{Fig:1}.

The standard version of the  M3Y parametrization \cite{M3YReview} fixes the parameter values
$c_{\rm C}, d_{\rm C}$ as given in Tab. \ref{Tab:1}. Instead of the Pauli blocking  ${\tilde W}^{\rm Pauli}(n_B)$
used for potentials A and B, the potential C contains the M3Y exchange part
$V_{\rm ex}= -276(1-0.005\,\,Q_{\alpha}/A_{\alpha})\delta(s)$
which also accounts for the Pauli exclusion principle, as well as an additional repulsive interaction simulating the Pauli blocking \cite{Misicu2007}.
The additional repulsive interaction $v^{\rm Pauli}(s)=v_{\rm rep}(s)=470 \delta(s)$ MeV is folding with a relatively
 density profile characterized by diffuseness parameter $a_{\rm rep}=0.268$ fm fitted to the nuclear
incompressibility $K=220$ MeV \cite{Misicu2007}.
Bound state energy, preformation factor, and half-life are calculated as shown in  Tab. \ref{Tab:1}.

As clearly seen in Fig. \ref{Fig:1},
all potentials A, B, and C are dominated by the Coulomb repulsion for finite
distances $r \geq 15$ fm, and at large distances only the bound
state energy of the free $\alpha$ particle remains, $\lim_{r \to
\infty}W( r)= -28.3$ MeV. Below $r \approx 15$ fm, both the
attractive nuclear potential and repulsive Pauli blocking between
the $\alpha$-particle and the lead core become relevant. At a
critical distance $r_{\rm cluster} =7.4383$ fm  (where $n_B=0.02917$ fm$^{-3}$), the
$\alpha$ cluster is suddenly dissolved and the four nucleons added
to the core are implemented on top of the Fermi energy $\mu_4$
\cite{Po}.

\begin{figure}[th]
\includegraphics[width=0.4\textwidth]{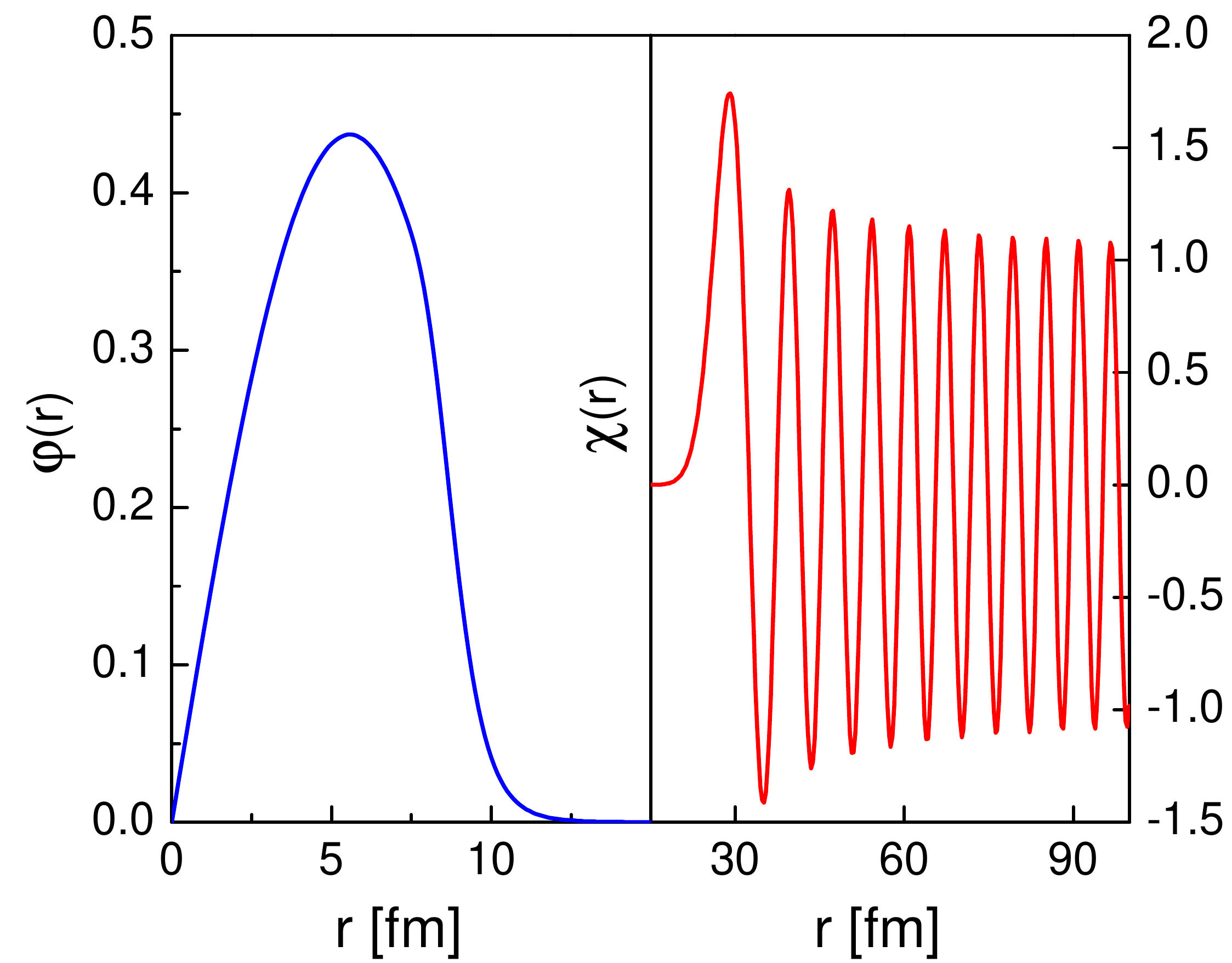}
\caption{The bound state wave function $\Phi(r)$ and the
scattering state wave function $\chi(r)$ calculated by separating
potential B into two parts based on the two-potential approach.
The separating point is chosen to be $r_{\rm sep}=15$ fm.}
 \label{Fig:2}
 \end{figure}

{\it Frequency (pre-exponential factor) $\nu$ and
exponential factor ${\cal T}$.} Using the two-potential approach of Gurvitz \cite{Gurvitz1988},
the effective c.o.m. potential $W({r})$ is separated into two
parts at $r_{\rm sep}=15$ fm  (the precise choice of the separating point will almost not
affect the final results). By solving the corresponding c.o.m.
Schr\"odinger equations, both the bound state wave function
$\Phi(r)$ and the scattering state wave function $\chi(r)$ are
calculated. We show both $\Phi(r)$ and $\chi(r)$ obtained from
potential B in Fig.~\ref{Fig:2}. The c.o.m. wave function $\Phi(r)$ exhibits an
approximately linear increase up to the critical distance
$r_{\rm cluster} =7.4383$ fm  (where $n_B=n_B^{\rm Mott}$) and then decreases.
As shown in \cite{Po}, the four-nucleon intrinsic wave function $\varphi^{\rm intr}({\bf s}_j,r)$
is nearly identical with the free $\alpha$-particle wave function in the region $r > r_{\rm cluster}$,
whereas for $r < r_{\rm cluster}$ the intrinsic wave function behaves like a product of
free nucleon wave functions so that the overlap with the free $\alpha$-particle wave function
is nearly zero.
The preformation probability of the $\alpha$ cluster is obtained
by integrating the $\Phi(r)$ from this critical point to
infinity  \cite{Po}:
\begin{eqnarray}
P_{\alpha}=\int_0^\infty  d^3r |\Phi(r)|^2 \Theta
\left[n_B^{\rm Mott}-n_B(r)\right]
\end{eqnarray}
The scattering state wave function $\chi(r)$ exhibits a strong
oscillating feature as a combination of regular and irregular
Coulomb functions. The decay width is then calculated by using the
values of $\Phi(r)$ and $\chi_k(r)$ at the
separation point. We choose $r_{\rm sep}=15$ fm \cite{Gurvitz1988}:
\begin{eqnarray}
\Gamma=\nu \times {\cal T}=\frac{4\hbar^2\alpha^2}{\mu k}|\Phi(r_{\rm sep})\chi_k(r_{\rm sep})|^2,
\end{eqnarray}
where $\mu = A_{\alpha}A_d/(A_{\alpha}+A_d)$, $\alpha
=\sqrt{2\mu(V(r_{\rm sep})-E_{\rm tunnel})}/{\hbar}$, $k=\sqrt{2\mu
E_{\rm tunnel}}/{\hbar}$, $A_d$ is the mass number of the lead core,
and the decay half-life is related to
the preformation probability and decay width by
$T_{1/2}=\hbar\ln2/(P_{\alpha}\Gamma)$.

{ \it Results.} In Table I, the details of the calculated preformation probability
and decay half-life of $^{212}$Po are presented. All potentials A, B, and C
are designed (parameter values $c, d$ for A, B, or exchange potential for C)
so that the experimental bound state energy
$E_{\rm tunnel}=Q_{\alpha}-28.3=8.954-28.3=-19.346$ MeV is reproduced. If the Thomas-Fermi
condition $E_{\rm tunnel}=\mu_4$ is fixed (potential A), the calculated half-life $T_{1/2}$
is too short. The measured decay
half-life $T_{1/2}$=$2.99\times 10^{-7}$ s is used to design potential B.
The corresponding Fermi energy of potential B is
$\mu_\alpha=-19.771$ MeV and the $\alpha$-cluster preformation
factor is $P_{\alpha}=0.142$. It is emphasized that the
preformation factor and decay half-life of $^{212}$Po are
consistently computed in a microscopic way, but the potential
B was chosen (parameter values $c_{\rm B}=11032.08$ MeV fm and $d_{\rm B}=3415.56$ MeV fm) to fit these empirical data.
The parametrization of potential C used is taken from the literature \cite{Misicu2007}, modified by fitting the
experimental binding energy. The calculated half-life $T_{1/2}$
is below the experimental value.
The $\alpha$-cluster
preformation factor, which is the most difficult part in the
$\alpha$-decay theory, is now well constrained by the experimental
data.

\begin{figure}[th]
\includegraphics[width=0.5\textwidth]{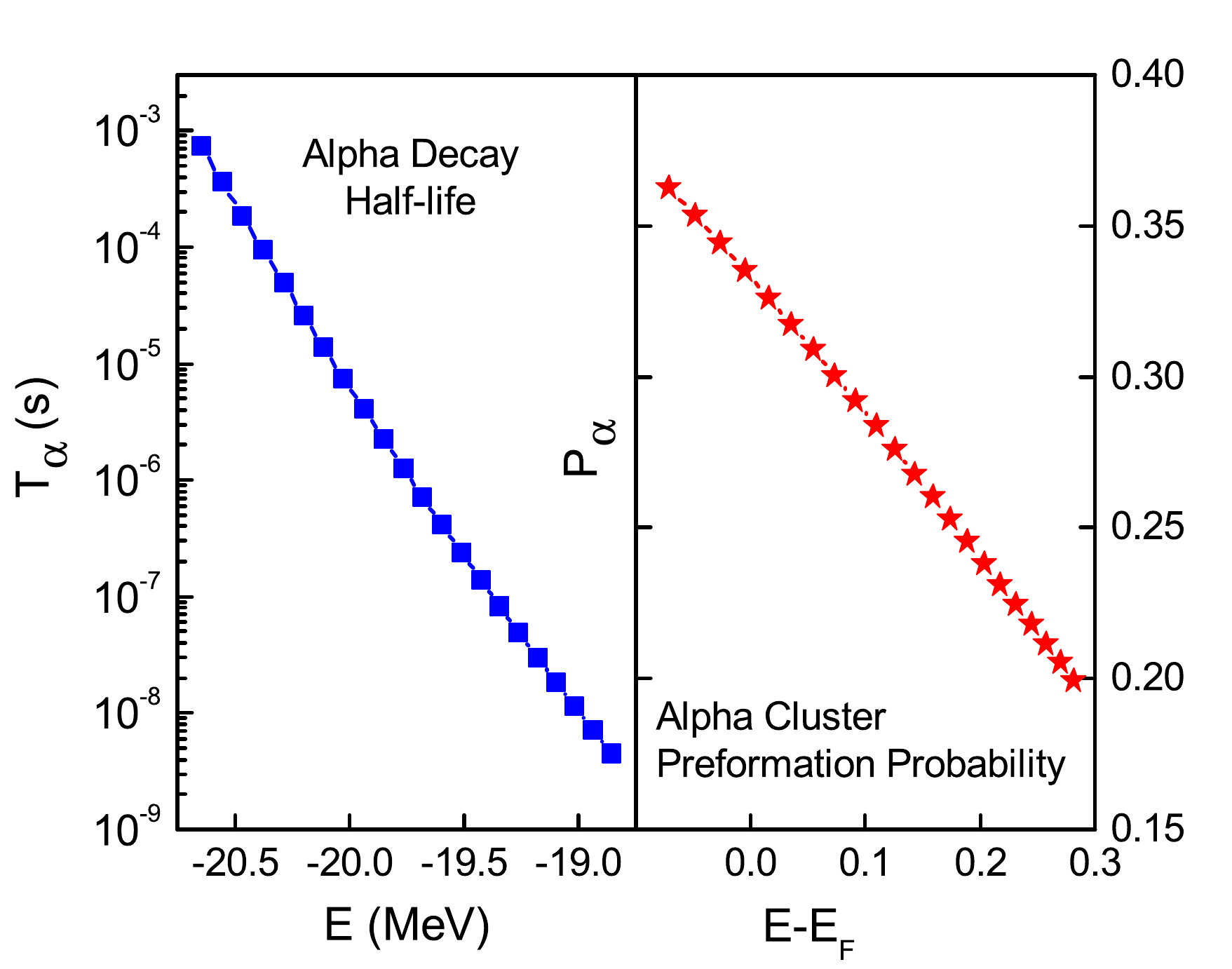}
\caption{The variation of the decay half-life $T_{1/2}$
with the energy eigenvalue  $E_{\rm tunnel}$ (left panel) and the variation of
preformation factor with the difference between the the energy
eigenvalue  $E_{\rm tunnel}$ and the Fermi energy $\mu_4$ (right panel).}
 \label{Fig:3}
 \end{figure}

By varying the strength parameter $v^0$ of the additional repulsive interaction in potential C $v_{\rm rep}(s)=v^0\delta(s)$ MeV while keeping other parameters fixed, it is observed that the decay half-life is mainly determined by
the energy eigenvalue $E_{\rm tunnel}$ (or decay energy). This is consistent with
previous $\alpha$-decay calculations. To show this behavior more
clearly, the correlation between the energy eigenvalue $E_{\rm tunnel}$ and the
decay half-life $T_{1/2}$ is given in the left panel of
Fig.~\ref{Fig:3} with various parameter values. Comparing potentials A, B, and C, it is
also observed that the $\alpha$ cluster preformation factor
depends closely on the difference between the the energy
eigenvalue $E_{\rm tunnel}$ and the Fermi energy $\mu_4$. A systematic
dependence of the preformation factor on $E_{\rm tunnel}-\mu_4$ is
shown in the right panel of Fig. \ref{Fig:3}.

\begin{table*}[htb]
\caption{The calculated preformation probability and decay
half-life of $^{212}$Po using different sets of effective c.o.m.
potentials.}\label{Tab:1}
\begin{tabular}{|l|ccc|ccc|cc|}
\hline \hline Potential &$c$&$d$ & exchange term& $E_{\rm tunnel}$   & Fermi energy  &
$E_{\rm tunnel}-\mu_4$ & Preform. factor  & Decay half-life
 \\
& [MeV fm]& [MeV fm]& & [MeV] & $\mu_4$[MeV] & [MeV] &  $P_\alpha$& $T_{1/2}$[s]\\
\hline
A      & 13866.30  &4090.51 &${ W}^{\rm Pauli}(n_B)$&  -19.346  & -19.346   & 0     &  0.367 &  2.91$\times 10^{-8}$\\
\hline
B      & 11032.08 & 3415.56 &${ W}^{\rm Pauli}(n_B)$&  -19.346  & -19.771   & 0.425 &  0.142 &  2.99$\times 10^{-7}$\\
\hline
C         &7999& 2134 & $V_{\rm ex}+V_{\rm Pauli}   $&-19.346  & -19.490  & 0.144 &  0.268 &  1.11$\times 10^{-7}$\\
\hline
\hline
\end{tabular}
\end{table*}

{\it Discussion.} We neglected gradient terms so that our approach is close to the
local-density approximation. We have to remember that within a
rigorous approach the c.m. potential $W({\bf r},{\bf r}')$ is
non-local. A full treatment of the inhomogeneous case relevant for
finite nuclei, which includes the gradient terms and the non-local
potentials, is a future goal, presently not in reach.

In addition, the relation (1) for the preformation probability is
only a simple approximation. An improved approach for the intrinsic
wave function $\varphi^{\rm intr}({\bf s}_j; {\bf R})$, with a smooth
behavior at $r_{\rm cluster}$ to replace the
step function $\Theta\left[n_B^{\rm Mott}-n_B(r)\right]$, will modify the
result for $P_\alpha$.

The core is described by an uncorrelated quasiparticle model,
the Thomas-Fermi model or nuclear shell model with a Fermi energy.
Also pairing correlations can be introduced.
It is an important question to improve the description of the core, allowing also for correlations.
This would affect the questions about the potential and the wave function for $r < r_{\rm cluster}$ where
a constant Fermi energy or chemical potential $\mu_4$ was considered.
Instead, the c.o.m. potential $W(r)$ and consequently the wave function $\Phi (r)$ will depend on $r$
in a more complex way also for $r < r_{\rm cluster}$. Presently, there are different attempts (FMD, AMD, etc.)
to include few-nucleon correlations to calculate light nuclei.
The treatment of heavier nuclei is not feasible with present computer capabilities.

The approach is inspired by the THSR wave function concept that
has been successfully applied to light nuclei. Shell model
calculations are improved by including four-particle
($\alpha$-like) correlations that are of relevance when the matter
density becomes low. A closer relation of the calculation
presented here to the THSR calculations is of great interest, see
the calculations for $^{20}$Ne \cite{Bo,Bo2}. Related calculations
are performed in Ref. \cite{Horiuchi}. The comparison with THSR
calculations would lead to a better understanding of the
microscopic calculations, in particular the c.o.m. potential, the
c.o.m. wave function, and the preformation factor.

(Note that the position of the chemical potential determines the
preformation probability. If $\mu_4<E_{\rm tunnel}$, the
preformation probability becomes smaller. If $\mu_4> E_{\rm
tunnel}$, the preformation probability becomes larger. In the
latter case, the states at the Fermi surface are also correlated.
This is the case, e.g., for the Hoyle state where all nucleons are
found in correlated states.)

\end{document}